\documentstyle[12pt]{article}

\def\bs{\begin{subequations}}
\def\es{\end{subequations}}

\catcode`\@=11

\newtoks\@stequation
\def\subequations{\refstepcounter{equation}
  \edef\@savedequation{\the\c@equation}%
  \@stequation=\expandafter{\theequation}%   %only want \theequation
  \edef\@savedtheequation{\the\@stequation}% % expanded once
  \edef\oldtheequation{\theequation}%
  \setcounter{equation}{0}%
  \def\theequation{\oldtheequation\alph{equation}}}

\def\endsubequations{\setcounter{equation}{\@savedequation}%
  \@stequation=\expandafter{\@savedtheequation}%
  \edef\theequation{\the\@stequation}\global\@ignoretrue}
\catcode`\@=12

\catcode`\@=11
\def\vereq#1#2{\lower3pt\vbox{\baselineskip1.5pt \lineskip1.5pt
\ialign{$\m@th#1\hfill##\hfil$\crcr#2\crcr\sim\crcr}}}
\catcode`\@=12

\makeatletter%  allow access to internal LaTeX commands
        \renewcommand{\theequation}{\thesection.\arabic{equation}}%
        \@addtoreset{equation}{section}%
\makeatother%  turn off access to internal LaTeX commands

\renewcommand{\thefootnote}{\fnsymbol{footnote}}

\def\la{\langle 0 \mid}
\def\ra{\mid 0 \rangle}

\begin{document}
\begin{titlepage}

July 18, 1999   
\begin{center}        \hfill   \\
            \hfill     \\
                                \hfill   \\

\vskip .25in

{\large \bf Computations in Large N Matrix Mechanics \\}

\vskip 0.3in

Charles Schwartz\footnote{E-mail: schwartz@physics.berkeley.edu}

\vskip 0.15in

{\em Department of Physics,
     University of California\\
     Berkeley, California 94720}
        
\end{center}

\vskip .3in

\vfill

\begin{abstract}
The algebraic formulation of Large N matrix mechanics recently developed 
by Halpern and Schwartz leads to a practical method of numerical 
computation for both action and Hamiltonian problems.  The new 
technique posits a boundary condition on the planar connected parts
 $X_{w}$, namely that they should decrease rapidly with increasing order.  
 This leads to algebraic/variational schemes of computation 
which show remarkably rapid convergence in 
numerical tests on some many-matrix models.  The method 
 allows the calculation of all moments of the ground state, in a 
sequence of approximations, and  excited states can be determined
as well.  There are two unexpected findings: a large d expansion and a new 
selection rule for certain types of interaction.
\end{abstract}

\vfill

\end{titlepage}

\renewcommand{\thefootnote}{\arabic{footnote}}
\setcounter{footnote}{0}
\renewcommand{\thepage}{\arabic{page}}
\setcounter{page}{1}

\section{Introduction}%1

Large N matrix mechanics \cite{Halp} differs from ordinary quantum mechanics in 
that the canonical commutator
\begin{equation}
i[p,q] = I,
\end{equation}
in the one-matrix case, is replaced by the relation
\begin{equation}
i[\pi,\phi] = \ra \la
\end{equation}
where $\ra$ is the ground state in the reduced Hilbert space. The 
original matrix-valued coordinates $\phi_{rs}, \;\; r,s = 1\ldots N$ 
 are represented by the single operator $\phi$ in this reduced 
 Hilbert space.~\cite{Bard} 
 
  The solution of the one-matrix Large N Hamiltonian 
problem with an arbitrary potential $V(\phi)$ was given some years 
ago~\cite{us1}; and only a couple of two-matrix problems in the 
action formalism have previously been solved.~\cite{Stau,Zinn}

The many-matrix problem involves several \emph{noncommuting} 
operators $\phi_{m}$ and their conjugate momenta.  Following 
 Halpern and Schwartz~\cite{us2}, this system is described at equal 
 times 
by a symmetric free algebra which involves a pair (tilde and untilde) 
for each hermitian operator
\bs
\begin{eqnarray}
[\tilde{\phi}_{m},\phi_{n}] = [\tilde{\pi}_{m},\pi_{n}] = 0, 
\;\;\;\;\;\; m,n=1\ldots d  \\ 
i[\tilde{\pi}_{m},\phi_{n}] = i[\pi_{m},\tilde{\phi}_{n}] \label{i} 
\label{b}
=\delta_{mn} \ra \la \\ 
\tilde{\phi}_{m} \ra = \phi_{m} \ra, \;\;\;\;\;\;
\tilde{\pi}_{m} \ra = \pi_{m} \ra \label{c}
\end{eqnarray}
\es
and the ground state energy is given by
\begin{equation}
E_{0} = N^{2} \la \frac{1}{2}\sum_{m=1}^{d}\pi_{m}\pi_{m} + V(\phi) \ra 
\end{equation}
where $(\phi)$ refers to the set of operators $\{\phi_{m}\}$. We 
shall use the summation convention in what follows.

In ordinary quantum mechanics systems of several interacting bodies 
are most commonly attacked from the Schrodinger equation in coordinate 
space, using the direct product basis $\mid q_{1},q_{2},\ldots q_{d} 
\rangle$. That approach is not available in the Large N reduced 
Hilbert space  
because of the noncommutativity of the operators $\phi_{m}$.  A basis 
of states in this reduced  space may be written as
\begin{equation}
\mid w \rangle \equiv \phi^{w} \ra
\end{equation}
where we use the ``word'' notation for ordered products of operators
\begin{equation}
\phi^{w} = \phi_{m_{1}} \phi_{m_{2}} \ldots \phi_{m_{n}}, 
\;\;\;\;\;\; w = m_{1}m_{2}\ldots m_{n}, \;\;\;\;\;\; m_{i} = 1 
\ldots d 
\end{equation}
and we write $[w]=n$ for the length of the word $w$.  See App.~A for a 
collection of relevant definitions and formulas.

The new approximation technique presented in this paper lies close to the 
Heisenberg (matrix) formulation rather than  the Schrodinger (wavefunction) 
formulation and makes use of the set of polynomials $T_{w}(\phi)$ 
introduced in Ref.~\cite{us3}.
\bs
\begin{eqnarray}
(1-\beta_{m}\phi_{m} + X(\beta))^{-1} \!\!&=&\!\! \sum_{w} 
\beta^{w}T_{w}(\phi) \\
X(\beta)=\sum_{w}\beta^{w} X_{w}, \;\;\;\;\;\; X_{0} \!\!&=&\!\! 0, 
\;\;\;\;\;\; \la T_{w}(\phi)\ra = \delta_{w,0} \label {e}
\end{eqnarray}
\es
where the $\beta_{m}$ are a dummy set of (noncommuting) parameters 
and the numbers $X_{w}$ were identified as the planar connected 
parts defined in earlier diagrammatic studies.~\cite{Cvit}  
 Various properties of these $X_{w}$ are given in App.~A, 
including their relation to the ordinary moments
 $Z_{w} \equiv \la \phi^{w} \ra$ of the ground state.

The core idea of the present work is to truncate the set of these $X$'s
\begin{equation}
set \;\;\; X_{w} = 0 \;\;\;\; for \;\; all \;\; [w] > n
\end{equation}
and solve the (now finite) set of algebraic equations, calling 
 this the ``n-th order approximation''. Then increase n, step 
by step, and see whether the numerical results appear to 
converge.
This is an intuitive/experimental approach for now, since we have no 
mathematical proof that this method should work.

With even a small number of the $X$'s determined, one can approximate 
all the moments of the ground state and the accuracy of 
these results increases systematically as one proceeds to higher 
orders of approximation.  The excited states of a Hamiltonian system 
are also amenable to this method.

The recent algebraic developments by Halpern and 
Schwartz~\cite{us2,us3} provide a wealth of formal definitions and 
relations for many-matrix problems, unifying the study of both action and 
Hamiltonian systems.  These start  
with the definitions of generalized creation and annihilation 
operators in the reduced Hilbert space,
\bs
\begin{eqnarray}
\pi_{m} \ra =iF_{m}(\phi) \ra, \;\;\;&&\;\;\; \la \pi_{m} 
= -i \la F_{m}(\phi)  \label{ag} \\ 
B_{m} = F_{m}(\phi) + i\pi_{m}, \;\;&\;\;&\;\; B_{m} \ra =
 \la B^{\dagger}_{m} = 0 \label{ae} \\ 
B_{m} B^{\dagger}_{n} \!\!&=&\!\! E_{mn}(\phi) \label{af} \\
E_{mn}(\phi) \ra \!\!&=&\!\! 2i[\tilde{\pi}_{n},F_{m}(\phi)] \ra \label{ah}
\end{eqnarray}
\es
which is the Interacting Cuntz Algebra. (In the case of 
non-interacting harmonic oscillators, we have $E_{mn} \propto \delta_{mn}$ 
and Eqs.~(\ref{ae}, \ref{af}) reduce to the original Cuntz algebra.)

In the practical work of this paper there is a basic distinction 
between the two types of problems. For action problems 
we start out knowing the functions $F_{m}(\phi)$ explicitly and this 
lets us work directly with the algebraic equations for the connected 
parts $X_{w}$ derived in Ref.~\cite{us3} (see Sec.~5).  For 
Hamiltonian problems we do not know $F_{m}(\phi)$ beforehand and so part of the 
method presented here involves a constructive representation of these 
operators, for which task we use the polynomials $T_{w}(\phi)$ (see 
Sec.~6).

In Sec.~2 we test the idea on a simple example: a one-matrix action 
problem and in Sec.~3 we try to give some understanding of why this 
method apparently works well. Counting of the variables in many-matrix 
problems and making use of symmetry to keep things manageable is 
discussed in Sec~4, followed in Sec.~5 by some algebraic results for 
a model action problem with d interacting matrices.
The plan of attack for many-matrix Hamiltonian problems is set out in 
Sec.~6 and  numerical results for a set of model potentials are presented in 
Sec.~7. We note not only the extremely rapid convergence found in 
these examples but also an unexpected selection rule. 
Section~8 presents  more details of this computational program;  
and a related method for calculating excited states is given in Sec.~9. 
Several appendices discuss further details and possible extensions of this work.

\section{First Test: One-Matrix Action Problem}%2

We start with a simple problem: a one-matrix action at Large N.  As 
given in Ref.~\cite{us3} for the quartic action ($F=\phi^{3}$), we have the 
following equation for the connected parts $X_{n}$:
\begin{equation}
X(X+1)^{2} -\beta^{2}X_{2} - \frac{1}{2}\beta^{4} = 0, \;\;\;\;\;\; X = 
\sum_{n>0}\beta^{n}X_{n} 
\end{equation}
which leads to the recursion formula
\begin{equation}
X_{n} = \frac{1}{2}\delta_{n,4} - \sum_{p=2}^{n-2}X_{p}(\sum_{q=2}^{n-p-2}
X_{q}X_{n-p-q} + 2X_{n-p}), \;\;\;\;\;\; n = 4,6,\ldots\;. \label{a}
\end{equation}
For one-matrix problems we replace the word label $w$ by $n=[w]$. 
We can compare this with the Schwinger-Dyson equations for the 
ordinary moments $Z_{n}= \la \phi^{n}\ra$, which may be written as
\begin{equation}
2Z_{n+4} = \sum_{m=0}^{n}Z_{m}Z_{n-m}, \;\;\;\;\;\; Z_{0}=1
\end{equation}
and only even n enter because of the parity symmetry in this 
problem.   If we have the value 
of $X_{2}=Z_{2}$ (which we know from other analysis to be $(2/3)^{3/2}$), 
then we can 
compute all the higher ones from these recursion formulas.  Table 
1 shows some numerical results and we see that the ratio $X_{n}/Z_{n}$
 decreases fairly rapidly as n increases.

\vskip 1cm

\begin{tabular}{|c||c|c|c|c|c|}
\multicolumn{6}{l}{Table 1. $X_{n}$ and $Z_{n}$ for $F=\phi^{3}$ action 
problem} \\
 \hline  
 n & 2 & 4 & 6 & 10 & 20  \\ \hline
 $X_{n}$ & .544331 & -.0925926 & .0403208 &  .0143736 &  -.00311591\\
$Z_{n}$ & .544331 & .500000 & .544331 & .816497 & 3.95996\\
 $X_{n}/Z_{n}$ & 1.00000 & -.185185 & .074074 & .017604 & .000787\\ \hline
\end{tabular}

\vskip 1cm
\nopagebreak 

Now we want to turn this process around and calculate the value of 
$X_{2}$ from the recursion formula (\ref{a}) using the idea that $X_{n}$ 
should decrease rapidly at large n  --  a sort of boundary condition.
That is, we consider $X_{2}$ as an unknown parameter and then search 
for that value that will allow us to truncate the  equations (\ref{a}) with 
$X_{n+2} = 0$; and then we step up the value of n and repeat the process.
Table 2 contains the results of this computation and we see that the 
residual error at each level of approximation decreases quite rapidly as we 
increase n.

\vskip 1cm

\begin{tabular}{|c||c|c|c|c|c|}
\multicolumn{6}{l}{Table 2. Compute $X_{2}$ by truncation: $X_{n+2}=0$} \\
 \hline  
 n+2 &  4 & 6 & 8 & 10 & 20  \\ \hline
Approx. $X_{2}$ & .500000 & .534522 & .541429 &  .543344 &  .544321\\
Error &-.044331 & -.009809 & -.002902 & -.000987 & -.000010 \\
  \hline
\end{tabular}

\vskip 1cm
\nopagebreak 

We view this as a sort of eigenvalue problem for the connected parts 
$X_{n}$ and recognize a certain similarity here with the familiar  
procedure for numerical integration of the one-dimensional Schrodinger
 equation in some 
given potential.  While that other problem involves a continuous 
variable $\psi(x)$ obeying a linear (differential) equation our current problem 
involves a discrete set $X_{n}$ obeying a nonlinear (algebraic) equation.

\section{Why Should this Method Work?}%3 

To understand what is going on here it may help to consider the 
ordinary moments
\begin{equation}
Z_{n} = \la \phi^{n}\ra = \int dq\; \rho(q)\;q^{n}
\end{equation}
for a one-matrix problem.  These $Z_{n}$, for a typical
 ground state density $\rho(q)$, are a  
rather monotonous sequence of numbers.  The infinite set of coupled 
equations for these moments (Schwinger-Dyson equations in one 
language) contains all the information about the ground state; but one 
would not try to truncate this infinite system of equations by setting 
the $Z_{n}$ equal to zero after some cutoff $n=n^{*}$.

(In earlier work~\cite{Blan} on moment equations for the one- and two-body 
Schrodinger equation, the asymptotic behavior of these moments as $n 
\rightarrow \infty$ was 
 inferred from the differential equation for the wavefunction and 
this allowed a backward iteration procedure.)

Now, by contrast, observe the definition of the planar connected parts, 
again for the one-matrix problem:
\begin{equation}
X_{n+1} = \la \phi T_{n}(\phi) \ra = \int dq \;\rho(q)\; q \;T_{n}(q) \label{aa}
\end{equation}
where the polynomials $T_{n}$ have the property
\begin{equation}
\la T_{n}(\phi) \ra = 0 \;\;\;\;\;\; n>0. \label{ab}
\end{equation}
Clearly, the $X_{n}$ are just an algebraic combination of the $Z_{n}$.
But Eq.~(\ref{ab}) tells us that the polynomials $T_{n}$ are 
oscillatory within the domain of integration; and this suggests 
that the $X_{n}$, given by (\ref{aa}), can be thought of as something 
like the Fourier coefficients  of the density $\rho(q)$.
Therefore, if the ground state is reasonably smooth and the polynomials 
$T_{n}$ are reasonably ``appropriate'', then we would expect that the 
higher Fourier coefficients (the $X_{n}$) could decrease rapidly. 
This is the motivation to try a truncation scheme on the $X$'s.

A further advantage of the $X$'s is that they are directly sensitive 
to the interactions in many-matrix problems. In Ref.~\cite{us3} 
 it was shown that in many-matrix problems without interactions, 
the $X_{w}$ vanish if there is any mixing of letters in the word $w$.

Once one has determined, approximately, even a small number of the
$X$'s, this allows one to give approximate values for \emph{all} 
of the $Z$'s in any one- or many-matrix problem by use of the general algebraic 
relation (\ref{ad}) between the generating functions for these two 
sets of parameters.

With these encouraging results, we go on to study the  problems 
of many matrices in  Large N action and Hamiltonian systems.

\section{Many Matrices - Counting the Variables}%4

With d matrices, the number of words of length n is $d^{n}$ 
and this number grows very rapidly. If we have some symmetries in the 
action or the Hamiltonian, then we can reduce the number of 
independent variables  
$X_{w}$ that we have to handle at each level of approximation. In this 
paper we consider model problems with the following 
invariance properties of the  ground state $\ra$. \\ 
\emph{Parity Symmetry}: Change the sign of $\phi_{m}$ (and $\pi_{m}$) 
 for any m. \\ 
\emph{Permutation Symmetry}: Make any permutation among the d labels $m,n,\ldots 
\;$.\\ 
In addition, there is the general invariance of the $X_{w}$ (as of 
the Trace operation in the unreduced space) under a cyclic 
permutation of the letters in the word $w$.

With these conditions, the number of independent $X_{w}$'s is greatly reduced
 - to what we shall call a set of ``basic 
 words'' at each level n - as shown in  Table 3. 
 
\vskip 1cm

\begin{tabular}{|r||c|c|c|c|}
\multicolumn{5}{l}{Table 3. Count of $d^{n}\rightarrow$ basic words} \\
 \hline  
 n & d=2 & d=3 & d=5 & d=9 \\ \hline
 2 & 4$\rightarrow$1 & 9$\rightarrow$1 & 25$\rightarrow$1 & 
 81$\rightarrow$1 \\
 4 & 16$\rightarrow$3 & 81$\rightarrow$3 & 625$\rightarrow$3 & 
 6561$\rightarrow$3 \\
 6 & 64$\rightarrow$4 & 729$\rightarrow$9 & 15625$\rightarrow$9 & 
 531441$\rightarrow$9 \\ 
 8 & 256$\rightarrow$12 & 6561$\rightarrow$41 & 390625$\rightarrow$59 &   \\
 10 & 1024$\rightarrow$28 &59049$\rightarrow$257 &  &  \\ 
 12 & 4096$\rightarrow$94 & & & \\ \hline
\end{tabular}

\vskip 1cm
\nopagebreak 

At each level of approximation (signified by the maximum word length n) we shall
 deal with 
a number of basic words (the dimension D of our parameter space).  
From Table 3 we read off these dimensions: for d=2, D=1,4,8,20,48,\ldots; for 
d=3, D=1,4,13,54,\ldots; for d=5, D=1,4,13,72,\ldots; for d=9, D=1,4,13,\ldots .
The first task of the computer program is to make a table of all 
$d^{n}$ words at each n, identify each word with an equivalence class 
according to the symmetries described above and assign one member of 
each class as a basic word $w_{i},\;\; i=1\ldots D$.

\section{Many-Matrix Action Problems}%5

\subsection{General algebraic machinery}%5.1

For action problems, we have the dual basis system of equations 
derived by Halpern and Schwartz~\cite{us3}:
\bs
\begin{eqnarray}
B_{m}^{\dagger} = G_{m}(\phi) - E_{mn}(\phi) \bar{B}_{n},&& 
\;\;\;\;\;\; \phi_{m} = \bar{B}_{m}(1+\bar{X}(B^{\dagger})) 
\label{ai} \\ 
\bar{X}(B^{\dagger}) = \sum_{w}X_{\bar{w}}B^{\dagger\;w} \!\!\!&=&\!\!\! 
\sum_{w} X_{w} B^{\dagger \;\bar{w}}.
\end{eqnarray}
\es
Here, the operators $\bar{B}_{m},\; B^{\dagger}_{m}$ obey the simple 
Cuntz algebra 
\begin{equation}
\bar{B}_{m} B^{\dagger}_{n} = \delta_{mn}
\end{equation}
and the role of these operators is to generate an infinite set of 
coupled algebraic equations for the connected parts $X_{w}$, as will 
be shown by example below.  The functions $G_{m}=2F_{m}$ and 
$E_{mn}$, defined earlier in (\ref{ag}, \ref{ah}), are immediately known 
once we specify the action $S$.  Then we shall proceed with the 
sequence of truncation approximations, generalizing the one-matrix 
example of Sec.~2.

\subsection{A model problem}%5.2

We take for our model problem here the d-matrix action
\begin{equation}
S = -\frac{1}{4N} \sum_{m,n=1}^{d}\; Tr([\phi_{m},\phi_{n}])^{2} 
\label{t}
\end{equation}
in the unreduced Hilbert space.  This gives us the reduced operators,
\bs
\begin{eqnarray}
&&G_{m}(\phi) = \sum_{n\neq m} (\phi_{m}\phi_{n}\phi_{n} + \phi_{n}\phi_{n}\phi_{m} 
- 2\phi_{n}\phi_{m}\phi_{n})  \label{ak} \\
&&E_{mm}(\phi) = \sum_{n \neq m}(\phi_{n}\phi_{n}+ X_{nn})  \\
&&E_{m \neq n}(\phi) = \phi_{m}\phi_{n} -2 \phi_{n}\phi_{m}
\end{eqnarray}
\es 
where we note that this $S$ has the symmetries mentioned in the 
previous section and this leads to the simplifications 
$X_{m} = 0, \;\; X_{mn} = \delta_{mn}X_{11}$.

The equations (\ref{ai}) now look like this
\begin{eqnarray} \label{am}
&&B^{\dagger}_{m} =  \sum_{n \neq m =1}^{d}\{ \nonumber \\ 
&&(\bar{B}_{n}\bar{B}_{n}\bar{B}_{m}\bar{X} + \bar{B}_{m}(\bar{B}_{n}
\bar{B}_{n}\bar{X} 
-X_{11}) -2 \bar{B}_{n}\bar{B}_{m}\bar{B}_{n}\bar{X}) + \nonumber \\ 
&&(\bar{B}_{n}\bar{X}\bar{B}_{n}\bar{B}_{m}\bar{X} + 
\bar{B}_{m}\bar{X}\bar{B}_{n}\bar{B}_{n}\bar{X} 
-2 \bar{B}_{n}\bar{X}\bar{B}_{m}\bar{B}_{n}\bar{X}) + \nonumber \\ 
&&(\bar{B}_{n}\bar{B}_{n}\bar{X}\bar{B}_{m}\bar{X} + 
\bar{B}_{m}\bar{B}_{n}\bar{X}\bar{B}_{n}\bar{X} 
-2 \bar{B}_{n}\bar{B}_{m}\bar{X}\bar{B}_{n}\bar{X}) + \nonumber \\
&&(\bar{B}_{n}\bar{X}\bar{B}_{n}\bar{X}\bar{B}_{m}\bar{X} + 
\bar{B}_{m}\bar{X}\bar{B}_{n}\bar{X}\bar{B}_{n}\bar{X} 
-2 \bar{B}_{n}\bar{X}\bar{B}_{m}\bar{X}\bar{B}_{n}\bar{X}) \}.
\end{eqnarray}

This system of equations is equivalent to the Schwinger-Dyson set 
of equations but it is packaged to emphasize the role of the $X$'s and 
it leads directly to our sequence of approximations. The first 
line of terms in (\ref{am}) has only one $\bar{X}$ and its first few terms are
\begin{equation}
(X_{mnnp} + X_{nnmp}-2X_{nmnp})B^{\dagger}_{p} + 
(X_{mnnpqr} + 
X_{nnmpqr}-2X_{nmnpqr})B^{\dagger}_{r}B^{\dagger}_{q}B^{\dagger}_{p}
\end{equation}
where the usual constraint on the sum ($n \neq m$) is understood.  The 
second and third lines have two $\bar{X}$'s and their first few terms are
\begin{eqnarray}
&&2X_{11}X_{mp} B^{\dagger}_{p} + (2X_{11}X_{mpqr}+ X_{np}X_{mnqr} +  
X_{mp}X_{nnqr} - \nonumber \\ 
&&2X_{np}X_{nmqr} + X_{nnpq}X_{mr}+X_{nmpq}X_{nr} -  
2X_{mnpq}X_{nr})B^{\dagger}_{r}B^{\dagger}_{q}B^{\dagger}_{p}
\end{eqnarray}
and the fourth line, with three $\bar{X}$'s, starts off as
\begin{equation}
(X_{np}X_{nq}X_{mr}+X_{mp}X_{nq}X_{nr}-2X_{np}X_{mq}X_{nr})
B^{\dagger}_{r}B^{\dagger}_{q}B^{\dagger}_{p}.
\end{equation}

Collecting the linear terms in $B^{\dagger}$ gives us the equation 
\begin{equation}
1 = 2 (d-1)(X_{1122}-X_{1212}+X_{11}^{2}) \label{aj}
\end{equation}
where we have used the symmetry properties to list the basic words:  
$(11)$ at $n=2$ and $(1111), (1122), (1212)$ at $n=4$.  This equation is 
exact  and leads to 
our lowest (second) order approximation: we set all $X$'s with word length 
greater than 2 equal to zero and we get 
\begin{equation}
X_{11} \simeq 1/\sqrt{2(d-1)} .
\end{equation}

Next, we collect the cubic terms in $B^{\dagger}$.  For our fourth order 
approximation we drop all $X_{w}$'s with $[w] >4$:
\begin{eqnarray}
0&=&X_{11}\{(2d-2+\epsilon_{mp}+\epsilon_{mr})X_{mpqr}-2\epsilon_{mp}X_{mqrp} 
-2\epsilon_{mr}X_{mrpq} + \nonumber \\ 
&&(\delta_{mp}\delta_{qr}+\delta_{mr}\delta_{pq})[(d-1)X_{1122} + 
X_{1111} - X_{mmqq}]\}+ \nonumber \\
&&X_{11}^{3}[\delta_{pq}\delta_{mr}\epsilon_{mp} +\delta_{qr}\delta_{mp}
\epsilon_{mq} -2 \delta_{pr}\delta_{mq}\epsilon_{mp}]
\end{eqnarray}
where $\epsilon_{pq} = 1-\delta_{pq}$.  These equations are now 
evaluated for varying choices of the labels $m,p,q,r$, which must be 
paired. We find
\bs \label{al}
\begin{eqnarray}
X_{1111}+X_{1122} = 0 \;\;\; for \; m=p=q=r \\ 
X_{1111}+ 3(d-1)X_{1122}-2X_{1212} + X_{11}^{2} = 0 \;\;\; for\; m=p 
\neq q=r \\ 
2X_{1122} - dX_{1212} + X_{11}^{2}=0 \;\;\; for \; m=q\neq p=r .
\end{eqnarray}
\es
The solution of this set of equations (for $d\neq 2$) is
\begin{equation}
X_{1111} = -X_{1122} =\frac{1}{3}X_{1212} = \frac{1}{3d+2}X_{11}^{2} 
\end{equation}
and, putting these results back into (\ref{aj}), we find the fourth order 
approximation for $X_{11}$
\begin{equation}
X_{11} \simeq [2(d-1)(1-\frac{4}{3d+2})]^{-\frac{1}{2}}.
\end{equation}

For $d=2$ the equations (\ref{al}) are indeterminate; but for this case 
 a scaling argument 
leads to the conclusion that the system is not bounded.

It was very pleasing to find, in the fourth order calculation above, 
that the number of independent equations was 
just equal to the number of unknowns and we found a unique solution. 
Will this circumstance continue at higher orders of approximation? 
I have no general answer.

One should  program a computer to carry the above sequence 
of approximations to higher order; only algebraic work is required at 
each step.  I have not done this yet, giving priority to the more 
difficult Hamiltonian problems, reported in Sec.~6.

\subsection{A large d expansion}%5.3

From the result above one is led to speculate that this truncation sequence of 
approximations may be related to a ``large d'' expansion.
The algebraic calculations described above have been carried out to the 
sixth order, with 9 equations in 9 unknowns, and solved in the
approximation that $d>>1$. This leads to the following result:
\begin{equation}
(X_{11})^{-2} = 2(d-1)[1 - \frac{4}{3d+2} - \frac{185}{81d^{2}} + 
O(d^{-3})]. \label{u}
\end{equation}

We do not have a systematic theory of such a large d 
approximation but the following crude attempt may be instructive. 
Look back at the formula for $G_{m}$, Eq.~(\ref{ak}), 
and replace the operator pair $\phi_{n}\phi_{n}$ by its ground 
state average, which is $X_{11}$.  This butchered $G_{m}$ is then
\begin{equation}
G_{m} \sim 2\omega \phi_{m}, \;\;\;\;\;\; \omega = (d-1)X_{11}
\end{equation}
which is the formula for a system of noninteracting harmonic oscillators.
The oscillator result $X_{mn} = \delta_{mn}/(2\omega)$ then gives 
immediately the leading term in (\ref{u}).  The higher order terms in $1/d$ 
are then expected to come from a perturbation theory expansion about 
this oscillator approximation.   Also, if one looks at the 
  computer results for the Hamiltonian problems (Sec.~7) 
one may discern a suggestion of more rapid convergence for larger values of d.

\section{Many-Matrix Hamiltonian Problems}%6 

\subsection{Choosing the model problems}%6.1

We shall study the Hamiltonians for d bosonic matrices, 
 given in the unreduced Hilbert space as
\begin{equation}
H = \frac{1}{2}\sum_{m=1}^{d}Tr(\pi_{m}\pi_{m}) + N 
Tr(V(\frac{\phi}{\sqrt N}))
\end{equation}
with the following choices of the potential:
\bs
\begin{eqnarray}
V_{1}(\phi) &=& \frac{1}{4} \sum_{m=1}^{d} \phi_{m}^{4} \\ 
V_{2}(\phi) &=& \frac{1}{4}(\sum_{m=1}^{d} \phi_{m}^{2})^{2}\label{y} \\ 
V_{3}(\phi) &=& \frac{1}{4}\sum_{m<n = 1}^{d} 
\phi_{m}^{2}\phi_{n}^{2}\label{z} \\ 
V_{4}(\phi) &=&-\frac{1}{8}\sum_{m<n = 1}^{d} [\phi_{m},\phi_{n}]^{2}
\end{eqnarray}
\es
or, if desired, any linear combination of them.
The first potential, which is just the non-interacting case, is used for 
verification of the computational procedure.  The third and fourth 
potentials have ``flat directions'', which make them particularly 
interesting. (Will the calculations converge nicely, indicating a 
bound state, or will they not?)  All four potentials have the 
symmetries (parity and permutation) described in Sec.~4. 
 The additional 
SO(d) symmetry of $V_{2}$ and $V_{4}$ is not used at the outset but 
will be noted in the results.

The following subsections  outline  the method and
further details are given in Sec.~8 and in Apps.~A and B.

\subsection{Construction of $F_{m}(\phi)$}%6.2

A central construct of our previous work~\cite{us2,us3} is the reduced 
operator $F_{m}(\phi)$, defined in (\ref{ag}). 
We will represent this quantity by a finite linear expansion in the 
polynomials $T_{w}(\phi)$
\begin{equation}
F_{m}(\phi) = \sum_{w} R^{(m)}_{w} T_{w}(\phi) \label{f}
\end{equation}
at each level of approximation and then see how to determine the 
coefficients $R$. (See Subsec~8.1 for more details.). 

For any reduced operator $A$ which depends on the $\phi$'s one has the 
identity
\begin{equation}
2 \la  A(\phi) F_{m}(\phi)\ra  = \la i[\tilde{\pi}_{m},A(\phi)]\ra  
\end{equation}
which is proved using the definition (\ref{ag}) and (\ref{c}).
 Choosing $A=T_{w'}$ and using the 
formulas (\ref{d}) and (\ref{e}) this gives
\begin{equation}
\la T_{w'}(\phi) F_{m}(\phi) \ra = \frac{1}{2} \delta_{w',m}
\end{equation}
for any word $w'$.  We impose these relations on the approximate 
expansion (\ref{f}) and obtain
\begin{equation}
\sum_{w} K_{w',w} R^{(m)}_{w} = \frac{1}{2} \delta_{w',m} \label {l}
\end{equation}
where
\begin{equation}
K_{w',w} \equiv \la  T_{w'}(\phi) T_{w}(\phi)\ra. \label{k}
\end{equation}
This matrix $K$ is numerically evaluated in terms of the $X$'s, as 
detailed  
in App.~B, and then we determine the expansion coefficients $R$ from a 
straightforward matrix inversion calculation. Of course, we make this 
a square (and positive) matrix, as detailed in Eqs.~(\ref{g}, \ref{h}). 
 This completes the 
first part of the fitting problem, which we would term the kinematic 
part since it assures that we are doing our best, at any given level 
of approximation, to represent the basic commutator algebra (\ref{i}).

Now we turn to the second part, which involves the dynamics of any 
particular Hamiltonian.

\subsection{Minimizing the Energy}%6.3

The kinetic energy of the ground state can be expressed as 
\begin{eqnarray}
 K.E./ N^{2} &=& \frac{1}{2} \la \pi_{m}\pi_{m} \ra = \frac{1}{2} \la  
F_{m} F_{m}\ra \nonumber \\ 
&=& \frac{1}{4} \la i[\tilde{\pi}_{m},F_{m}]\ra 
= \frac{1}{4} R^{(m)}_{m} = \frac{d}{4} R^{(1)}_{1}
\end{eqnarray}
using the methods and results of the previous subsection.

The potential energy of the ground state is expressed directly in 
terms of the $X$'s using (\ref{j}):
\bs
\begin{eqnarray}
\la \phi_{m}^{4} \ra &=& X_{1111}+ 2X_{11}^{2} \\
\la \phi_{m}^{2} \phi_{n}^{2} \ra &=& X_{1122}+X_{11}^{2}, \;\;\; m 
\neq n \\
\la \phi_{m}\phi_{n}\phi_{m}\phi_{n} \ra &=& X_{1212}, \;\;\; m 
\neq n 
\end{eqnarray}
\es
where we have used the specified symmetries to write these formulas in terms of 
the four basic words at the second and fourth orders.

We  program the computer to evaluate the ground state 
energy $E=E_{0}/N^{2}$ at the n-th order approximation with any 
assigned  
numerical values for the quantities $X_{w}$ for $[w] \leq n$.
The final step of this scheme is to vary this set of $X$'s so as to 
minimize  E.  This procedure is without 
mathematical justification; it just seems like the natural thing to do.

What is more, this part of the method is far from straightforward as a 
computational task because the energy E is a very nonlinear function 
of the many variables $X$.  In Subsec.~8.2 we describe the 
techniques  
used to search for this minimum.  The numerical results are presented 
next.

\section{Numerical Results}%7

The  Tables that follow give the outputs of the computations and are designed
 to show at a glance the convergence of the approximation 
scheme described above.

Table 4 shows the energy (E/d) calculated for the potential $V_{2}$, for 
several values of d and at several levels of approximation and Table 5 
gives the corresponding values of $X_{11}= \la \phi_{1}^{2}\ra$. 

\vskip 1cm

\begin{tabular}{|r|c||l|l|l|l|}
\multicolumn{6}{l}{Table 4. Calculated values of E/d for potential $V_{2}$} \\
 \hline  
 n & D & d=2 & d=3 & d=5 & d=9 \\ \hline
 2 & 1 & .429 & .472 & .5408 & .6412\\
 4 & 4 & .42672 & .47035 &.53921 & .64007\\
 6 & 8,13 & .426672 & .4703152 & .539189 & .640058\\ 
 8 & 20,54 & .42667093 & .47031461 & & \\ 
 10 & 48 & .426670885 & & & \\ \hline
\end{tabular}

\nopagebreak 

\vskip 1cm

\begin{tabular}{|r|c||l|l|l|l|}
\multicolumn{6}{l}{Table 5. Calculated values of $X_{11}$ for potential $V_{2}$} \\
 \hline  
 n & D & d=2 & d=3 & d=5 & d=9 \\ \hline
 2 & 1 & .437 & .397 & .347 & .292\\
 4 & 4 & .4428 & .4010 &.34912 & .29365\\
 6 & 8,13 & .443007 & .401106 & .349171 & .293667\\ 
 8 & 20,54 & .4430170 & .4011103 & & \\ 
 10 & 48 & .44301744 & & & \\ \hline
\end{tabular}

\vskip 1cm
\nopagebreak 

We note  how rapidly these numbers  converge as 
one goes down each column in the tables.  For each step increasing
the order of approximation, we see  \emph{one or two orders of 
magnitude} increase in accuracy, somewhat better for $E$ than for $X$. 
 Also, one sees in these tables that the 
first approximation (a `back of the envelope' computation) is accurate to about 
one percent.  Such is the power of the X.  For  comparison, Table 6 
presents results for the one-matrix problem, d=1 and $V_{1}$, computed 
by the same program.  We see that 
the results of the many-matrix computations (above) converge  about as 
rapidly as the one-matrix results, although the amount of work 
required to obtain the former is much greater.

\vskip 1cm

\begin{tabular}{|r|c||l|l|}
\multicolumn{4}{l}{Table 6. Computed results for the one-matrix 
problem: $V_{1}$} \\
 \hline  
 n & D & E & $X_{11}$ \\ \hline
 2 & 1 & .375 & .50 \\
 4 & 2 & .3717 & .5100 \\
 6 & 3 & .371638 & .51057 \\ 
 8 & 4 & .3716339 & .510611 \\ 
 10 & 5 & .37163373 & .5106136 \\ \hline
\end{tabular}

\vskip 1cm
\nopagebreak 

Table 7 gives the E/d results computed for the potential $V_{3}$ and 
one sees rapid convergence here as well.

\vskip 1cm

\begin{tabular}{|r|c||l|l|l|l|}
\multicolumn{6}{l}{Table 7. Calculated values of E/d for potential $V_{3}$} \\
 \hline  
 n & D & d=2 & d=3 & d=5 & d=9 \\ \hline
 2 & 1 & .236 & .298 & .375 & .4725\\
 4 & 4 & .2312 & .29470 &.373207 & .471358\\
 6 & 8,13 & .231036 & .294625 & .3731823 & .47134965\\ 
 8 & 20,54 & .2310258 & .29462242 & & \\ 
 10 & 48 & .23102504 & & & \\ \hline
\end{tabular}

\vskip 1cm
\nopagebreak

In Table 8 we see  the results for the potential $V_{4}$, which 
has the greatest amount of ``flat directions'' among our models.  Here 
the rate of convergence is noticeably slower than in the previous models, 
but still looks convincingly good. 

\vskip 1cm

\begin{tabular}{|r|c||l|l|l|l|}
\multicolumn{6}{l}{Table 8. Calculated values of E/d for potential $V_{4}$} \\
 \hline  
 n & D & d=2 & d=3 & d=5 & d=9 \\ \hline
 2 & 1 & .24 & .30 & .38 & .47\\
 4 & 4 & .224 & .289 &.370 & .4690\\
 6 & 8,13 & .2232 & .2890 & .36944 & .468940\\ 
 8 & 20,54,72 & .22299 & .28895 &.369431 & \\ 
 10 & 48 & .222964 & & & \\ \hline
\end{tabular}

\vskip 1cm
\nopagebreak 

Also, in the several tables above, one sees a suggestion of more rapid 
convergence for larger values of d; see the discussion of the large d 
expansion in Subsec.~5.3.

\vskip 1cm 

 In another experiment, we 
studied the one-matrix problem with potential
\begin{equation}
V(\phi) = \frac{1}{2} \phi^{2} - \frac{g}{4}\phi^{4}
\end{equation}
as the parameter $g$ approached the value $\sqrt{8}/3\pi$  where the bound 
state disappears.  The numerical  procedure searching to minimize the 
energy worked well 
until one approached very close to this critical value; then it 
failed dramatically.

\vskip 1cm

 Other $X_{w}$ values are also produced in these 
computations, albeit with a somewhat lesser accuracy.  Table 9 has some 
of these for the potential $V_{2}$.

\vskip 1cm

\begin{tabular}{|c||c|c|c|c|}
\multicolumn{5}{l}{Table 9. Computed values of some other $X_{w}$ for $V_{2}$} \\
 \hline  
$X_{w}$ & d=2 & d=3 & d=5 & d=9 \\ \hline
$X_{1111}$ & -.0132659 & -.0082358 & -.004201 & -.001798 \\ 
$X_{1122}$ & -.0066329 & -.0041179 & -.002101 & -.000899 \\ 
$X_{1212}$ & 0.0 & 0.0 & 0.0 & 0.0 \\ \hline
\end{tabular}

\vskip 1cm
\nopagebreak

\noindent If there is rotational symmetry in the ground state, one can derive 
the following relation among the fourth order $X$'s,
\begin{equation}
X_{1111} = 2 X_{1122} + X_{1212} \label{ac}
\end{equation}
and the data in Table 9 satisfy this relation, as does the 
corresponding data for the potential $V_{4}$, which is also 
rotationally invariant. 

\vskip 1cm 

There is another, unexpected, phenomenon seen in the data of Table 9: 
namely that $X_{1212} = 0$. An increasing number of other $X_{w}$'s also  
vanish when one looks  at higher orders. This result also appears for the 
potential $V_{3}$, but not for $V_{4}$. When a particular $X_{w}$ goes 
to zero, so does the corresponding coefficient $R_{w}$. The empirical 
rule is this: Write out the word $w$ and remove any pair of matching adjacent 
letters; repeat this process; the $X_{w}$ will vanish unless this process 
can reduce the original word to null.
 I do not have a full explanation 
for this newly discovered selection rule but it appears to be related to the 
fact that these potentials (see (\ref{y}) and (\ref{z})) involve only pairs 
($\phi_{m}\phi_{m}$) of each operator.  This new symmetry 
 is particular to Large N matrix mechanics with its 
noncommuting coordinate operators; it  would not arise in ordinary quantum 
mechanics.

\vskip 1cm 

From an experimental (numerical) perspective, but lacking any formal proof, 
it appears that these types of large N problems are now solvable.  It will be 
important for others to repeat this work independently in order to 
verify these results.

\section{Details of the Computational Program}%8

\subsection{The Full $F_{m}$}%8.1

The expression (\ref{f}) for $F_{m}(\phi)$ needs to be refined. The 
motivation for what follows comes from App.~E in Ref.~\cite{us2} where 
 the ground state wavefunction (the action) is modeled and one sees the 
consequent structure of $F_{m}(\phi)$.

Corresponding to each basic word $w_{i}$ we want to have a group of 
terms (in the $T_{w}(\phi)$) with a common coefficient $R^{(m)}_{i}$:
\begin{equation}
F_{m}(\phi) = \sum_{i=1}^{D} R^{(m)}_{i} F_{m,i}(\phi).
\end{equation}
For the first stage in this construction we define
\begin{equation}
\partial_{m} T_{w}(\phi) \equiv \sum_{w \sim mw'} T_{w'}(\phi)
\end{equation}
which, one can show, will guarantee that the flatness condition~\cite{us2}
\begin{equation}
[\tilde{\pi}_{m},F_{n}(\phi)] - [\pi_{n},\widetilde{F_{m}(\phi)}]= 0
\end{equation}
is satisfied.

For the second stage we take all permutations among the $m=1 \ldots d$ 
letters that occur in the basic words $w_{i}$.
\begin{equation}
F_{m,i}(\phi) = \frac{1}{c(w_{i})\;(d-1)!}\;\partial_{m}\;\sum_{perm's} 
 Permute \;T_{w_{i}}(\phi)
\end{equation}
where the constant $c(w)$, the number of subcycles in 
the word $w$, is defined in App.~A.  The normalization constants used 
above are convenient but not essential.

Now we construct the matrix elements
\begin{equation}
t_{i,j} = \la F^{\dagger}_{m,i}(\phi) F_{m,j}(\phi)\ra  \;\;\; (no\; sum) 
\label{g}
\end{equation}
where these are linear combinations of the $K_{w,w'}$ defined in 
(\ref{k}) and Eq.(\ref{l}) is replaced by
\begin{equation}
\sum_{j=1}^{D}t_{i,j}\; R^{(m)}_{j} = \frac{1}{2} \delta_{i,1}, 
\label{h}
\;\;\;\;\;\; i=1\ldots D \; .
\end{equation}
In order to save computing time in evaluating each $t_{i,j}$ it is important to 
find and to count repeated evaluations of the same $K$ elements.  I am 
not sure that I have done this job completely in my program.

\subsection{Searching}%8.2

The hardest part of this program is searching for the minimum energy in the 
parameter space of the basic word  connected parts: $X_{w_{i}} \equiv x_{i},
 \; i=1\ldots D$.  The first method used  fits a quadratic function to
E(x) evaluated at D(D+1)/2 nearby points and then finds the 
extremum:
\bs
\begin{equation}
b_{i} = E(x_{i}+\delta) -E(x), \;\;\;\;\;\;
a_{i,j} = E(x_{i}+\delta,x_{j}+\delta) -E(x) - b_{i}-b_{j} \label{m}
\end{equation}
\begin{equation}
\sum_{j=1}^{D}\;a_{i,j}\; v_{j} = b_{i} - \frac{1}{2} a_{i,i}, \;\;\;\;\;\; 
x'_{i} = x_{i} - \delta v_{i}.
\end{equation}
\es
If one is close enough to the minimum, iterating this procedure should 
converge rapidly.  For most of the data presented in Sec.~7 this 
method worked, although I am sure that  more sophisticated techniques 
could have been more efficient. For the largest size computation 
carried out (d=5, n=8, D=72) the time for each evaluation of the energy 
was about one minute and each iteration of this search procedure took 
about 44 hours on a common desktop microcomputer.

Sometimes, however, this approach failed.  For the potential $V_{4}$, 
beyond the sixth order calculation (for d=2 and d=3) this method 
diverged or led to impossible output (see the next subsection).  What 
succeeded in those cases was a second method: start by solving the 
numerical problem for some other potential (like $V_{3}$ where the first
search  method 
worked well) and then gradually change a coupling constant $g$ inserted 
into the potential and solve again, repeating in small steps until one arrives 
at the desired result.  At each new step one can start efficiently with a sort of 
perturbation theory
\begin{equation}
\sum_{j=1}^{D}a_{i,j} \Delta_{j} =\delta^{2}\;\frac{\partial^{2}E(x)}{\partial 
g\;\partial x_{i}}, \;\;\;\;\;\; x_{i}' = x_{i} - (\delta g) \Delta_{i}
\end{equation}
which involves the matrix $a_{i,j}$ (\ref{m}) which one has already 
calculated at the previous step.

Just because the numerical search appears to converge is no proof 
that we have found the correct solution.  In work on the potential $V_{4}$ 
for $d=2$ we had some results at the sixth order (D=8) which first 
appeared well converged by the first searching method;
 but a later check on the rotational 
symmetry (\ref{ac}) showed that this was a false solution.  Repeating this 
calculation using the second search method described above led to 
satisfactory results. The fact that the false energy value was off  only 
in the fifth decimal place stands as a cautionary note on this 
new numerical technique.

Another numerical searching procedure  is suggested 
by the algebraic work in Sec.~5.  One could vary only the 
subset of $X_{w}$'s with $[w]=n^{*}$, keeping all others fixed; then 
cycle through the choices of n*.

It should be repeated that this is all experimental work that is in 
need of  sound mathematical justification and guidance.  The 
multidimensional energy surface $E(x)$ is a very complicated nonlinear 
function of the parameters $x$.  In fact, there are singularities 
which may lie not far away from the desired minimum.  One can see the 
simplest example of this situation in the 2x2 matrix equation (\ref{h}) 
for the d=1 case.

\subsection{Constraints}%8.3

The quantities $X_{w}$ cannot be regarded as completely independent 
variables. For example, in the one-matrix case one has 
\begin{equation}
\la (\phi^{2} - <\phi^{2}>)^{2} \ra \geq 0
\end{equation}
which leads to the inequality $X_{4} \geq -(X_{2})^{2}$.

Using the general Schwarz inequality, we can write
\begin{equation}
|\la T_{w}T_{w'}\ra|^{2} \leq \la T_{\bar{w}}T_{w}\ra \la 
T_{\bar{w'}}T_{w'}\ra
\end{equation}
for all words $w$ and $w'$. This implies many constraints upon the 
allowed values of the $X$ parameters as we search to minimize the 
energy.  It is unclear how best to implement these constraints; in the 
computations reported here I only checked that the matrix (\ref{g}) satisfied
\begin{equation}
|t_{i,j}|^{2} \leq t_{i,i}\; t_{j,j},\;\;\;\;\;\; t_{i,i} > 
0\;\;\;\forall i,j 
\end{equation}
at each evaluation. A failure of this test signals that the search has 
strayed into forbidden territory.

An entirely different sort of constraint comes from the use of a 
purely real (rather than complex) representation for the $\phi$ operators. 
  This implies that we should have $X_{w} = X^{*}_{w} = X_{\bar{w}}$.  
With the extensive symmetry of the problems studied here many of these 
constraints are automatic; but at the 10th order for $d=2$ and at the 
8th order for $d>2$, one finds some basic words that do not satisfy
$\bar{w} \approx w$.  Rather than imposing this constraint, we 
are satisfied to find that this equality comes out in the numerical 
results.

\section{Excited States}%9

After the ground state problem is solved, we consider excited 
(adjoint) states in the reduced Hilbert space:
\begin{equation}
H \ra = E_{0} \ra, \;\;\;\;\;\;H \mid E \rangle = E \mid E \rangle
\end{equation}
where it should be remembered that we do not know the form of the 
reduced Hamiltonian $H$ \cite{us2} but only that it generates time 
translations.  With the postulate
\begin{equation}
\mid E \rangle = U \ra
\end{equation}
for some operator $U$ we find the identity
\begin{equation}
(E-E_{0}) \la U^{\dagger}U \ra = -i \la U^{\dagger} \dot{U} \ra. 
\label{o}
\end{equation}

Now we make the construction, as with $F$ before,
\begin{equation}
U = \sum_{w}r_{w} T_{w}(\phi), \;\;\;\;\;\; U^{\dagger} = \sum_{w} r^{*}_{w}
T_{\bar{w}}(\phi)
\end{equation}
and we have, using (\ref{n}), 
\begin{equation}
\dot{U} = \sum_{w} r_{w} \sum_{w=w_{1}mw_{2}} T_{w_{1}} \pi_{m} 
T_{w_{2}}
\end{equation}
where the $r_{w}$ are as yet undetermined constants.

We can now write (\ref{o}) as
\begin{equation}
E-E_{0} =( \sum_{w,w'}r^{*}_{w} L_{\bar{w},w'}r_{w'}) / 
(\sum_{w,w'}r^{*}_{w} K_{\bar{w},w'}r_{w'}) \label{p}
\end{equation}
where the matrix $K_{w,w'}$ was defined earlier and from (\ref{x}) we 
have
\begin{equation}
L_{w,w'}\equiv -i\la T_{w}\dot{T}_{w'}\ra  = \frac{1}{2}\sum_{w=umv} \sum_{w'=u'mv'}K_{u,v'} K_{u',v}.
\end{equation}
Finally, vary the coefficients $r$ to find stationary values of 
(\ref{p}) and we get a 
traditional linear matrix  problem, where $E-E_{0}$ is an eigenvalue 
of the matrix $L$ with respect to the metric matrix $K$.

The evaluation of the matrix $K$ and thus also of $L$ is done entirely 
in terms of the $X_{w}$'s, which were already solved with the ground state 
problem.  Thus (although I have not done any explicit numerical 
calculations for excited states) the complete spectrum of $H$ can be calculated.
The lowest order approximation, $U=T_{m}(\phi)$, gives $E_{m}-E_{0} 
= 1/(2X_{mm})$.

\vspace{1cm}

\noindent {\bf ACKNOWLEDGEMENTS}

I am grateful to M.~B.~Halpern for his advice on numerous occasions 
and I also thank K.~Bardakci and M.~Rieffel for helpful conversations.

\vskip 1.0cm
\setcounter{equation}{0}
\def\theequation{A.\arabic{equation}}
\boldmath
\noindent{\bf Appendix A. Useful Formulas Old and New}%A
\unboldmath
\vskip 0.5cm

Further conventions on the word notation: \\
$w=0$ is the null word. \\
$w=m$ means that the word $w$ consists of a single letter m.\\
$w \sim w'$ means that the two words differ by at most a cyclic 
permutation of \\ \indent their letters. \\
$w \approx w'$ means that the two words are equivalent under some 
larger symmetry.
$w_{1}w_{2} = w_{3}$ means that the second word is appended to the 
first word and \\ \indent the result is the third word. \\
$w = umv$ means that the word $w$ is decomposed as indicated. \\ 
$\bar{w}$ is the word formed by reversing the sequence of letters in 
the word $w$.
$c(w)$, the number of subcycles in the word $w$, is defined as the 
largest integer \\ \indent k such that $w=u^{k}$ for any word $u$ with
  $[u]>0$.

\vskip 12pt

Basic relations among $T(\phi)$ and $X$ \cite{us3}:
\bs
\begin{eqnarray}
T_{mw} = \phi_{m}T_{w} \!\!\!&-&\!\!\! \sum_{w=w_{1}w_{2}} X_{mw_{1}} T_{w_{2}} 
\label {q} \\ 
T_{wm} = T_{w}\phi_{m} \!\!\!&-&\!\!\! \sum_{w=w_{1}w_{2}}T_{w_{1}} X_{w_{2}m} 
\label{r} \\ 
X_{mw} = \la \phi_{m}T_{w}\ra \!\!\!&=&\!\!\! \la T_{w}\phi_{m}\ra = X_{wm} \\
T_{w}^{\dagger} = T_{\bar{w}}, \;\;\;&&\;\;\; X_{w}^{*} = X_{\bar{w}} .
\end{eqnarray}
\es

Relation between $X$ and $Z$:
\begin{equation}
Z(j) = 1 + X(j Z(j)). \label{ad}
\end{equation}
Examples (for the case of parity symmetry, which means that each 
letter must appear an even number of times or else the $Z$ and $X$ vanish):
\bs
\begin{eqnarray}
&&Z_{mn} = \la \phi_{m}\phi_{n}\ra = \delta_{mn}X_{mm} \\
&&Z_{mnpq} = Z_{npqm} = \left\{\begin{array} {l}
X_{mmmm}+2 X_{mm}^{2}\;\;\; if\;\; m=n=p=q \\
X_{mmpp} + X_{mm}X_{pp} \;\;\; if\;\; m=n \neq p=q \\
X_{mnmn} \;\;\; if\;\;p=m \neq n=q .\end{array} \right. \label{j}
\end{eqnarray}
\es
For one-matrix problems the label $w$ is replaced by $n=[w]$. For 
systems with parity selection rule:
\bs
\begin{eqnarray}
&&T_{0} = 1, \;\;\;\;\;\;T_{1} = \phi, \;\;\;\;\;\; T_{2} = \phi^{2} 
-X_{2}, \;\;\;\;\;\; X_{2} = <\phi^{2}> \\
&&T_{3} = \phi^{3}-2\phi X_{2}, \;\;\;\;\;\; T_{4} = \phi^{4}-3\phi^{2}
X_{2}-X_{4}+X_{2}^{2} \\
&&X_{4} = <\phi^{4}> - 2X_{2}^{2}, \;\;\;\;\;\; X_{6} = <\phi^{6}> - 
6X_{4}X_{2} -5X_{2}^{3} .
\end{eqnarray}
\es

Below are some new relations involving $T(\phi)$ that are used in the 
present work.  Start with the generating function
\begin{equation}
Y = 1/(1-\beta_{m}\phi_{m} + X(\beta)) = \sum_{w} 
\beta^{w}T_{w}(\phi)
\end{equation}
and calculate the commutator,
\begin{equation}
i[\tilde{\pi}_{m},Y] = Y \beta_{m} \ra \la Y.
\end{equation}
Now expand in  powers of $\beta$ and match 
terms to find:
\begin{equation}
i[\tilde{\pi}_{m},T_{w}(\phi)] = \sum_{w=w_{1}mw_{2}}T_{w_{1}}(\phi) 
\ra \la T_{w_{2}}(\phi). \label{d}
\end{equation}
The other version of this relation 
\begin{equation}
i[\pi_{m},\tilde{T}_{w}] = \sum_{w=w_{1}mw_{2}} \tilde{T}_{w_{2}}
\ra \la \tilde{T}_{w_{1}}
\end{equation}
comes from Eq.(D.11) in Ref.~\cite{us3}.
In a very similar way one gets the time derivative equation
\begin{equation}
\frac{d}{dt}T_{w}(\phi) = \sum_{w=w_{1}mw_{2}} T_{w_{1}} \pi_{m}T_{w_{2}}
 \label{n}
\end{equation}
where we have used $\frac{d}{dt}\phi_{m} = \pi_{m}$.
Combining the last two equations leads to
\begin{equation}
i \la T_{w'}\frac{d}{dt} T_{w} \ra = -\frac{1}{2}\sum_{w=umv}\;\sum_{w'=u'mv'} 
\la T_{u}T_{v'}\ra \la T_{u'}T_{v}\ra \label{x}
\end{equation}
which is surprisingly simple.

\vskip 1.0cm
\setcounter{equation}{0}
\def\theequation{B.\arabic{equation}}
\boldmath
\noindent{\bf Appendix B. Evaluating $<T_{w}T_{w'}>$}%B
\unboldmath
\vskip 0.5cm

We seek some recursive procedure for evaluation of the matrix elements
\begin{equation}
K_{w,w'} = \la T_{w}(\phi) T_{w'}(\phi) \ra = K_{w',w}
\end{equation}
in terms of the connected parts $X_{w}$.  Using equations (\ref{q}) and 
(\ref{r}) 
 it is relatively easy to find the following relations
\begin{equation}
K_{wm,w'} = K_{w,mw'}+ \sum_{w'=uv}X_{mu}K_{w,v}- \sum_{w=uv} X_{vm}
K_{u,w'}
\end{equation}
with the boundary counditions $K_{w,0}=K_{0,w}=\delta_{w,0}$.
This looks very nice as a recursive computer program but it turns out 
to be  expensive: the time required grows exponentially  as 
one increases the size of the words.  One could save time by building 
a table of all the $K$ matrix elements one might need, but that requires 
enormous amounts of space.

An alternative method is given by the following formula
\begin{equation}
K_{w,w'} = \sum_{w=uv} \sum_{w'=u'v'} K_{u,v'} X_{vu'}, \;\;\;\; 
[v] >0, \;\;\; [u']>0, \;\;\;\;\;\; K_{0,0}=1 \label{s}
\end{equation}
which may be derived by combining equation (\ref{q}) with the  
expansion
\begin{equation}
\phi_{m} = \sum_{w}X_{mw}G_{\bar{w}}(\phi)
\end{equation}
from Ref.~\cite{us3} and also using the identity
\begin{equation}
\la T_{w}T_{w'}G_{\bar{w}''}\ra = \sum_{w''=uv}\la T_{w_{1}} T_{w_{2}}\ra\; 
\delta_{w,uw_{1}} \; \delta_{w',w_{2}v}
\end{equation}
which is similar to the Ward Identities derived in App.~E of 
Ref.~\cite{us3}.

The program uses (\ref{s}) to build a small table of $K$'s each time one of 
them is called for and  the time for this grows as $n^{4}$ 
rather than exponentially.  Still, this is the main time consuming 
part of the computations.

\vskip 1.0cm
\setcounter{equation}{0}
\def\theequation{C.\arabic{equation}}
\boldmath
\noindent{\bf Appendix C. Some Alternative Computational Schemes}%C
\unboldmath
\vskip 0.5cm

One alternative scheme is to start out by fitting the quantity 
$E_{mn}(\phi)$ instead of $F_{m}(\phi)$,
\begin{equation}
E_{mn}(\phi) = \sum_{w} R^{(mn)}_{w} T_{w}(\phi).
\end{equation}
The definition (\ref{ah}) is
\begin{equation}
E_{mn}(\phi) \ra = 2i [\tilde{\pi}_{n},F_{m}(\phi)] \ra
\end{equation}
and using (\ref{d}), we find
\begin{equation}
R^{(mn)}_{w} = 2R^{(m)}_{wn}
\end{equation}
upon comparison with (\ref{f}).
Next, we  use the formal expansion from Ref.~\cite{us3}
\begin{equation}
(E^{-1})_{mn} = \sum_{w}X_{wmn} G_{\bar{w}}(\phi)
\end{equation}
to write the system of conditions
\bs
\begin{eqnarray}
&&\la T_{w'} (E_{mn}(E^{-1})_{np} - \delta_{m,p})\ra = 0 \\
&&\sum_{w,w''} X_{npw''}\la T_{w'}T_{w}G_{\bar{w}''} \ra R^{(m)}_{wn} = 
\frac{1}{2} \delta_{m,p}\delta_{w',0}
\end{eqnarray}
\es
and one can show, using (\ref{s}), that this reduces to equations identical to
(\ref{l}).  So this  method is not alternative at all.

A second alternative scheme does away with minimizing the energy and 
 works instead from the equations of motion:
\begin{equation}
\dot{\pi}_{m} \ra = i \dot{F}_{m}(\phi)\ra = - V'_{m}(\phi) \ra.  
\end{equation}
Using the representation (\ref{f}) for $F_{m}$, this leads us to a
new system of equations
\begin{equation}
-i \sum_{w} \la T_{w'} \dot{T}_{w}\ra R^{(m)}_{w} =  \la 
T_{w'}V'_{m}\ra \label{v}
\end{equation}
where the matrix elements on the left side are the quantities  
$L_{w',w}$ defined in Sec.~9. One now has \emph{two} sets of matrix 
equations - (\ref{l}) and (\ref{v}) - determining the same set of 
expansion coefficients: call the solutions $R$ and $R'$.  One would now 
seek a set 
of values for the parameters $X_{w_{i}}$ that would  make these two 
sets of solutions the same.  Computationally, the  way to do this would 
presumably be to minimize the error,
\begin{equation}
Error \; = \sum_{i}|R_{i} - R'_{i}|^{2}
\end{equation}
and this defines another nonlinear search procedure.
But what weight function ought optimally to be put into this error 
calculation? 

A third alternative is to use the monomials $\phi^{w}$ instead of the 
polynomials $T_{w}(\phi)$ as a basis for the fitting of  the operators 
$F_{m}$ or  $U$.  This leads to much simpler formulas for the matrix 
elements of $K$ and $L$, expressed in terms of the moments $Z_{w} = \la 
\phi^{w}\ra$.  Then one would use the relation (\ref{ad}) to evaluate each 
$Z_{w}$ in terms of the chosen set of parameters $X_{w}$.  I believe that this 
approach has drawbacks in both  speed and numerical accuracy; but it 
should be explored.

\vskip 1.0cm
\setcounter{equation}{0}
\def\theequation{D.\arabic{equation}}
\boldmath
\noindent{\bf Appendix D. Is this Method Useful in Ordinary QM?}%D
\unboldmath
\vskip 0.5cm

With the apparent success of this approximation method in Large 
N matrix mechanics, one goes back to ordinary quantum mechanics to see 
if we have a new useful calculational technique. 
The formalism developed in Ref.~\cite{us3} is easily modified to fit 
the standard commutation relation
\begin{equation}
i[p_{i},q_{j}] = \delta_{ij} I
\end{equation}
with the following construction:
\bs
\begin{eqnarray}
Y \!\!\!&=&\!\!\! e^{\beta_{i}q_{i} - X(\beta)} = \sum_{\mu}T_{\mu}(q) \\
X(\beta) \!\!\!&=&\!\!\! \sum_{\mu}C_{\mu} \beta^{\mu} X_{\mu} \\
Z(\beta) = \la e^{\beta_{i}q_{i}}\ra \!\!\!&=&\!\!\! \sum_{\mu}C_{\mu}\beta^{\mu}
\la q^{\mu}\ra.
\end{eqnarray}
\es
Here $\mu$ represents the unordered set of occupation numbers 
$\{\mu_{i}\}$ (remember that the $q_{i}$'s commute with one another now) and
\begin{equation}
C_{\mu} = 1/\prod_{i} (\mu_{i})!\;.
\end{equation}

In the simple one-matrix case we have
\bs
\begin{eqnarray}
&&Z(\beta) = \sum_{n=0}^{\infty}\beta^{n}Z_{n}/n!, \;\;\;\;\;\;Z_{n} = \la 
q^{n}\ra  = \int dq\; q^{n} \rho(q) \\
&&X(\beta)= \sum_{n=1}^{\infty}\beta^{n}X_{n}/n!  =  ln(Z(\beta))
\end{eqnarray}
\es
and  we want to test whether the ratio 
$X_{n}/Z_{n}$ decreases rapidly with n, as we saw for the Large N 
situation in Sec.~2.  For the case of a harmonic oscillator, we have 
the same result in both theories, namely $X_{n}$ vanishes for $n>2$.

One simple (non-oscillator) model that allows analytic calculations is 
a constant density $\rho(q)$ over some finite range of q.  Here we 
find that the ratio $X_{n}/Z_{n}$ decays rapidly with n 
 for the Large N 
situation but this ratio grows very rapidly for the ordinary quantum 
mechanics situation.

We have also applied the method of this paper to the  quantum 
mechanical nonlinear oscillator,
\begin{equation}
H = \frac{1}{2}p^{2}+ \frac{1}{4}q^{4}.\label{w}
\end{equation}  
Numerical results for the ground state are shown in Table 10. The 
convergence seen here is fairly good, although not as good as for the similar 
Large N problem, shown in Table 6. (The accuracy shown here is 
comparable to that obtained with conventional variational 
calculations of this Schrodinger equation, at the same levels of 
approximation.) 

\vskip 1cm

\begin{tabular}{|r|c||l|l|}
\multicolumn{4}{l}{Table 10. Results for the Schrodinger equation 
(\ref{w})}
\\
 \hline  
 n & D & E & $X_{11}$ \\ \hline
 2 & 1 & .429 & .437 \\
 4 & 2 & .4217 & .4525 \\
 6 & 3 & .4210 & .45512 \\ 
 8 & 4 & .42086 & .45571   \\ \hline
\end{tabular}

\vskip 1cm
\nopagebreak 

 It must be reported, however, that the results shown 
in Table 10 were not obtained easily.  The problem of nearby 
singularities in the energy surface, mentioned in Subsec.~8.2, was 
more severe in this ordinary quantum mechanics problem than in the 
Large N problems. For the calculations through $D=3$ I used the 
second searching method, starting from the harmonic oscillator and then moving 
gradually to the quartic potential in  steps of size 1/8.  For $D=4$ I had to 
decrease the step size to 1/16; and for $D=5$ I gave up after failing 
in the search procedure with step size 1/32.

In conclusion, I am still in doubt about the answer to the question 
posed in the heading of this appendix.

%****************************


\begin{thebibliography}{99}

\bibitem{Halp} 
M.~B.~Halpern, {\sl Nucl. Phys.}\/ {\bf B188}, 61 (1981).

\bibitem{Bard}
K.~Bardakci, {\sl Nucl. Phys.}\/ {\bf B178}, 263 (1981).

\bibitem{us1}
M.~B.~Halpern and C.~Schwartz, {\sl Phys. Rev.}\/ {\bf D24}, 2146 
(1981).

\bibitem{Stau} 
M.~Staudacher, {\sl Phys. Lett.}\/ {\bf B305}, 332 (1993).

\bibitem{Zinn}
V.~A.~Kazakov and P.~Zinn-Justin, ``Two-Matrix Model with ABAB 
Interaction'', hep-th/9808043.

\bibitem{us2} 
M.~B.~Halpern and C.~Schwartz, ``The Algebras of Large N Matrix 
Mechanics'', hep-th/9809197, to appear in IJMPA.  This paper provides 
a brief overview of related work in the field.

\bibitem{us3} 
M.~B.~Halpern and C.~Schwartz, ``Infinite Dimensional Free Algebra and 
the Forms of the Master Field'', hep-th/9903131, to appear in IJMPA.

\bibitem{Cvit}
P.~Cvitanovic, P.~G.~Lauwers, P.~N.~Scharbach, {\sl Nucl. Phys.}\/ 
{\bf B203}, 385 (1982).

\bibitem{Blan}
B.~Blankenbecler, T.~DeGrand and R.~I.~Sugar, {\sl Phys. Rev.} \/ 
{\bf D21}, 1055 (1980).

\end{thebibliography}
\end{document}